\documentclass{article}

\usepackage{arxiv}

\usepackage[utf8]{inputenc} % allow utf-8 input
\usepackage[T1]{fontenc}    % use 8-bit T1 fonts
\usepackage{hyperref}       % hyperlinks
\usepackage{url}            % simple URL typesetting
\usepackage{booktabs}       % professional-quality tables
\usepackage{amsfonts}       % blackboard math symbols
\usepackage{nicefrac}       % compact symbols for 1/2, etc.
\usepackage{microtype}      % microtypography
\usepackage{lipsum}
\usepackage{graphicx}
\usepackage{makecell}

\graphicspath{ {./images} }

\usepackage[caption=false,font=normalsize,labelfont=sf,textfont=sf]{subfig}
\usepackage{amsmath}
\usepackage{siunitx}
\usepackage[numbers]{natbib}

\title{Data-Driven Pseudo-spectral Full Waveform Inversion via Deep Neural Networks}

\author{
  Christopher Zerafa\\
  Department of Geosciences\\
  University of Malta\\
  Msida, Malta\\
  \texttt{christopher.zerafa.08@um.edu.mt} \\
   \And
   Pauline Galea \\
  Department of Geosciences\\
  University of Malta\\
  Msida, Malta\\
  \texttt{pauline.galea@um.edu.mt} \\
   \AND
   Cristiana Sebu \\
  Department of Geosciences\\
  University of Malta\\
  Msida, Malta\\
   \texttt{cristiana.sebu@um.edu.mt} 
}

\begin{document}
\maketitle
\begin{abstract}
  FWI seeks to achieve a high-resolution model of the subsurface through the application of multi-variate optimization to the seismic inverse problem. Although now a mature technology, FWI has limitations related to the choice of the appropriate solver for the forward problem in challenging environments requiring complex assumptions, and very wide angle and multi-azimuth data necessary for full reconstruction are often not available.
    
  Deep Learning techniques have emerged as excellent optimization frameworks. These exist between data and theory-guided methods. Data-driven methods do not impose a wave propagation model and are not exposed to modelling errors. On the contrary, deterministic models are governed by the laws of physics. 
  
  Application of seismic FWI has recently started to be investigated within Deep Learning. This has focussed on the time-domain approach, while the pseudo-spectral domain has not been yet explored. However, classical FWI experienced major breakthroughs when pseudo-spectral approaches were employed. This work addresses the lacuna that exists in incorporating the pseudo-spectral approach within Deep Learning. This has been done by re-formulating the pseudo-spectral FWI problem as a Deep Learning algorithm for a data-driven pseudo-spectral approach. A novel DNN framework is proposed. This is formulated theoretically, qualitatively assessed on synthetic data, applied to a two-dimensional Marmousi dataset and evaluated against deterministic and time-based approaches.
  
  Inversion of data-driven pseudo-spectral DNN was found to outperform classical FWI for deeper and over-thrust areas. This is due to the global approximator nature of the technique and hence not bound by forward-modelling physical constraints from ray-tracing.  
\end{abstract}

\section{Introduction}
Full waveform inversion (FWI) seeks to achieve a high-resolution model of the subsurface through the application of multivariate optimization to the seismic inverse problem \citep{Virieux2009}. Optimization theory is fundamental to FWI since the parameters of the system under investigation are reconstructed from indirect observations that are subject to a forward modelling process \citep{Tarantola2005}. Choice of the forward problem will impact the accuracy of the FWI result. Challenging environments require more complex assumptions to try and better explain the physical link between data and observations, with not necessarily improved levels of accuracies \citep{Morgan2013}. Secondly, the data being used to reconstruct the mapping of measurements for the ground-truth are not optimal. Very wide angle and multi-azimuth data is required to enable full reconstruction of the inverse problem \citep{Morgan2013}; which is not always available.

Recently, deep learning techniques have emerged as excellent models to solve for inverse problems \citep{Elshafiey1991,Adler2017a}. These deep learning based waveform inversion processes exist between data and theory guided methods \citep{Sun2019}. Data-driven methods do not impose a wave propagation model. Neural network weights are all trainable and require relatively exhaustive training datasets to invert properly \citep{Sun2019a}. Yet, due to the large number of degrees of freedom, they are not exposed to modelling errors as any conventional FWI algorithm \citep{Wu2018}.

Application to FWI has recently started to be investigated within the deep learning field and has so far been focused only on the time-domain approach \citep{Sun2019a}. However, classical FWI experienced pivotal breakthroughs via pseudo-spectral approaches \citep{Sirgue2006} which enabled the technique to go beyond academic experiments and be employed on real datasets\citep{Sirgue2009}. In this paper we want to investigate whether the same advantages apply when pseudo-spectral FWI is developed within data-driven Deep Neural Networks (DNN). To current knowledge, there is no prior work investigating the pseudo-spectral inversion within DNN frameworks. Hence, the main aim of this work was to investigate and develop a novel approach in the form of data-driven DNN pseudo-spectral FWI, compare it to traditional approaches and investigate its benefits and limitations. To this end, the following steps were followed:
\begin{enumerate}
    \item Re-casting FWI within a DNN framework for a data-driven learned inversion based DNN formulation. This is derived theoretically and assessed on synthetic data. 
    \item The results are validated against classical deterministic FWI.
    \item Analysing the limitations of the approach and discuss future potential developments.
\end{enumerate}

\section{Theoretical Background}
The forward problem in FWI is based on the wave equation describing particle wave motion. It is a second order, partial differential equation involving both time and space derivatives. The particle motion for an isotropic medium is given by:
\begin{equation}
\frac{1}{c(\mathbf{m})^2} \frac{\partial^2p(\mathbf{m},t)}{\partial t^2} - \nabla^2p(\mathbf{m},t) = s(\mathbf{m},t),
\end{equation}
where $p(\mathbf{m},t)$ is the pressure wave-field, $c(\mathbf{m})$ is the acoustic $p$-wave velocity and $s(\mathbf{m},t)$ is the source \citep{Igel2016}. To solve the wave equation numerically, it can be expressed as a linear operator. 
% Although the data $\mathbf{d}$ and model $\mathbf{m}$ are not linearly related, the wave-field $p(\mathbf{m},t)$ and the sources $s(\mathbf{m},t)$ are linearly related by the equation:
% \begin{equation}
% \mathbf{A}p(\mathbf{m},t) = s(\mathbf{m},t),
% \end{equation}
% where $p(\mathbf{m},t)$ is the pressure wave-field produced by a source $s(\mathbf{m})$ and $\mathbf{A}$ is the numerical implementation of the operator:
% \begin{equation}
% \frac{1}{c(\mathbf{m})^2} \frac{\partial^2}{\partial t^2} - \nabla^2,
% \end{equation}

A common technique employed within the forward modelling stage is to perform modelling in pseudo-spectral domain rather than the time domain. The most common domain is the Fourier domain \citep{Igel2016} and computational implementation is generally achieved via the Fast Fourier Transform (FFT) \citep{cooley1965algorithm}. 

% \begin{figure*}[b]
% 	\centering
% 	\includegraphics[width=0.95\textwidth]{fwi_workflow.png}
% 	\caption{Schematic of a FWI workflow solved as an iterative optimisation process.}
% 	\label{fig:algo_schem_fwi}
% \end{figure*}

After forward modelling the data in pseudo-spectral domain, the objective is to seek to minimize the difference between the observed data and the modelled data. The misfit between the two datasets is known as the objective- or cost-function $\emph{J}$. The most common cost function is given by the $L_2$-norm of the data residuals:
% \begin{equation}
% \emph{J}(\mathbf{m}) = \frac{1}{2}{\left|| d - F(\mathbf{m}) \right||}^2_D,
% \end{equation}
\begin{equation}
    \emph{J}(\mathbf{m}) = \frac{1}{2}\left[ {\left|| d - F(\mathbf{m}) \right||}^2_D + \lambda\left||\mathbf{m}\right||^2_M \right] ,
\end{equation}
where $D$ indicates the data domain given by $n_s$ sources and $n_r$ receivers, $M$ is the model domain, and $\lambda$ is a regularization parameter to introduce well-posedness. The misfit function $\emph{J}$ can be minimized with respect to the model parameters $d$ if the gradient is zero, namely:
\begin{equation}
\nabla\emph{J} = \frac{\partial\emph{J}}{\partial \mathbf{d}} = 0,
\end{equation}

% Minimising the misfit function is generally achieved via a linearised iterative optimisation scheme based on the Born approximation in scattering theory \citep{Born1980}. 
% The inversion algorithm starts with an initial estimate of the model $\mathbf{m}_0$. After the first pass via forward modelling, the model is updated by the model parameter perturbation $\Delta \mathbf{m}_0$. This newly updated model is then used to calculate the next update and the procedure continues iteratively until the computed model is close enough to the observations based on a residual threshold criterion.
At each iteration $k$, assuming small enough model perturbation and using Taylor Expansion up to second orders, the misfit function $\emph{J}(\mathbf{m}_k)$ is calculated from the previous iteration model $\mathbf{m}_{k-1}$ as:
%  by:
% \begin{equation} \label{eq:misfit_k-1}
% \emph{J}(\mathbf{m}_k) = \emph{J}(\mathbf{m}_{k-1} + \Delta \mathbf{m}_k),
% \end{equation}
\begin{equation} \label{eq:tay_exp}
    \emph{J}(\mathbf{m}_k) = \emph{J}(\mathbf{m}_{k-1}) 
		+ \delta \mathbf{m}^T_{k-1}
		\frac{\partial\emph{J}}{\partial\mathbf{m}_{k-1}}
		+ \frac{1}{2}
		\delta\mathbf{m}^{2T}_{k-1}
		\frac{\partial^2\emph{J}}{\partial\mathbf{m}^2_{k-1}},
\end{equation}

\section{Proposed FWI as a Data-Driven DNN}
Neural networks are a subset of tools in artificial intelligence which when applied to inverse problems can approximate the non-linear function of the inverse problem $F^{-1}:D\rightarrow M$. That is, using a neural network, a non-linear mapping can be learned to minimize:
\begin{equation}
\left||\mathbf{m} - g_{\theta}(\mathbf{d})\right||^2 ,
\end{equation}
where $\theta$ the large data set of pairs $(\mathbf{m}, \mathbf{d})$ used for the learning process \citep{Adler2017a}. 

\begin{figure}[h]
	\centering
	\includegraphics[width=0.5\textwidth,]{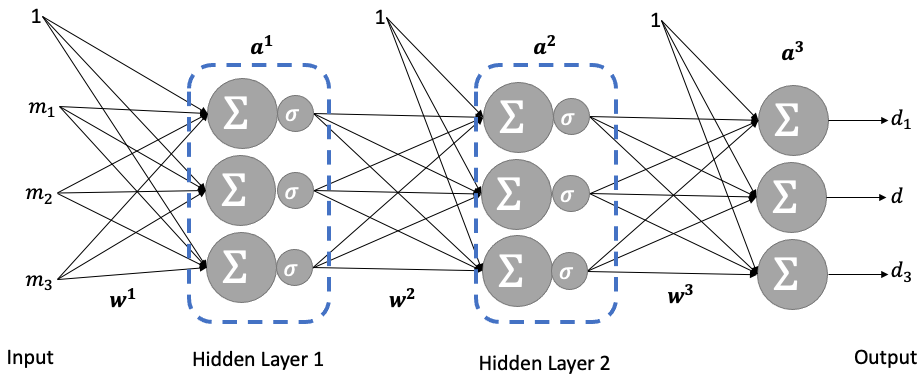}
	\caption{An example of a fully connected neural network with 2 hidden layers. All weights $w$ and bias $b$ are learned during the training phase. The $1$'s connected to each hidden layer represents bias nodes which help the neural network learn patterns by allowing the output of an activation function to be shifted. Adapted from \citep{Lucas2018}.}
	\label{fig:mlp_2_layer}
\end{figure}

The most elementary component in a neural network is a neuron. This receives input and sums the result to produce an output. For a given artificial neuron, consider $n$ inputs with signals $m$ and weights $w$. The output $d$ of the $k^{\text{th}}$ neuron from all input signals is given by:
\begin{equation}
d_k=\sigma\left( b+\sum_{j=0}^{m} w_{kj}m_{j} \right) ,
\end{equation}
where $\sigma$ is the activation function and $b$ is a bias term enabling the activation functions to shift about the origin.
When multiple neurons are combined together, they form a neural network. The architecture of a neural network refers to the number of neurons, their arrangement and their connectivity \citep{Sima2003}. Figure \ref{fig:mlp_2_layer} shows a neural network consisting of 2 hidden layers. The output of the unit in each layer is the result of the weighted sum of the input units, followed by a non-linear element-wise function. The weights between each units are learned as a result of a training procedure.

When training a DNN, the forward propagation through the hidden layers from input $\mathbf{m}$ to output $\mathbf{d}$ needs to be measured for its misfit. The most commonly used cost function is the Sum of Squared Errors, defined as:
\begin{equation}
\emph{J}(\mathbf{m}) = \frac{1}{2}\sum_{i=1}^{J}\left( \mathbf{m} - g_{\theta}(\mathbf{d}^{(i)}) \right)^{2},
\end{equation}
where $\mathbf{d}$ is the labelled true dataset, $\mathbf{d}^{(i)}$ is the output from the $i^{\text{th}}$ forward pass through the network and the summation is across all neurons in the network. The objective is to minimize the function $\emph{J}$ with respect to the weights $w$ of the neurons in the neural network. Employing the Chain Rule, the error signals for all neurons in the network can be recursively calculated throughout the network and the derivative of the cost function with respect to all the weights $w$ can be calculated. Training of the DNN is then achieved via a Gradient Descent algorithm, referred to as back-propagation training algorithm \citep{rumelhart1985learning}. 

% \subsection{Outline for solving FWI using DNN}
% Algorithm for training of a DNN for FWI is given in Algorithm \ref{algo:dnn} and a schematic is given in Figure \ref{fig:algo_schem_dnn}.
% \begin{algorithm}
% Setup a deep architecture for the neural network.\\
% Initialise the set of weights $w^l$  and biases $b^l$.\\
% Forward propagate through the network connections to calculate input sums and activation function for all neurons and layers.\\
% Calculate the error signal for the final layer $\delta^L$ by choosing an appropriate differentiable activation function.\\
% Back-propagate the errors $(\delta^l)$ for all neurons in layer $l$.\\
% Differentiate the cost function with respect to biases $\left(\frac{\partial\emph{J}}{\partial b^l} \right) $.\\
% Differentiate the cost function with respect to weights $\left( \frac{\partial\emph{J}}{\partial w^l}\right)$.\\
% Update weights $w^l$ via gradient descent.\\
% Recursively repeat from Step 3 until the desired convergence criterion is met.
% \caption{FWI as a data-driven DNN}
% \label{algo:dnn}
% \end{algorithm}

% \begin{figure*}[t]
% 	\centering
% 	\includegraphics[width=0.95\textwidth]{DNN_workflow.png}
% 	\caption{Schematic of a FWI workflow solved as learned optimisation process.}
% 	\label{fig:algo_schem_dnn}
% \end{figure*}

\section{Experimental Results}\label{sec:results_discussion}

\subsection{Train-Test Data Split}\label{sec:results_Train-Test_Data}
Learning the inversion from time to pseudo-spectral domain requires a training dataset which maps time to pseudo-spectral components and their respective velocity profile. A data generator was designed to create synthetic data for a 2000\si{ms} time window. The steps involved in the data generator are:
\begin{enumerate}
    \item Randomly create velocity profile $v_p$ for a 2000ms distance, with value ranging from 1450\si{ms^{-1}} and 4500\si{ms^{-1}}.
    %  The lower bound of 1400\si{ms^{-1}} was selected for the water column \citep{Cochrane1991} and upper bound of 4000\si{ms^{-1}} for velocity in porous and saturated sandstones \citep{Lee1996}. The assumption is made that limestones, carbonates and salt deposits are not present in the subsurface model since these would have velocity in excess of 4000\si{ms^{-1}}.
    \item Estimate the density $\rho$ using Gardner’s equation $\rho=\alpha v_p^{\beta}$ where $\alpha=0.31$ and $\beta=0.25$ are empirically derived constants~\citep{Gardner1974}.
    \item At each interface, calculate the Reflection Coefficient $\mathcal{R}=\frac{\rho_2v_{p_{2}} - \rho_1v_{p_{1}}}{\rho_2v_{p_{2}} + \rho_1v_{p_{1}}}$ where $\rho_i$ is density of medium $i$ and $v_p$ is the p-wave velocity of medium $i$.
	% \item For each medium, calculate the Acoustic Impedance $\mathcal{Z}=\rho v_p$.
    \item Define a wavelet $\mathcal{W}$. This was selected to be a Ricker wavelet at 10\si{Hz} \citep{Ryan1994}.
    %  The central frequency of 10\si{Hz} was chosen as a nominal value based on literature results to be representative of normal FWI conditions \citep{Morgan2013}. 
	\item The reflection coefficient time series and wavelet are convolved to produce the seismic trace $\mathcal{T}$.
	\item Fourier coefficients for magnitude $\mathcal{M}(\zeta)$ and phase $\mathcal{M}(\phi)$ are derived based on the FFT.
\end{enumerate} 

To exploit higher dimensionality, a secondary generator was designed to perform Continuous Wavelet Transforms (CWT). This was identical to the previous generator, expect that in Step (6), produce a continuous wavelet transform with sampling frequencies from 1-75\si{Hz}. 
% The different steps for these two generator flows are shown in Figure~\ref{fig:data_generators} for a sample velocity profile. 
These generators will be referred to as Generator 1 and Generator 2 respectively.

% \begin{figure}[h]
% 	\centering
% 	\includegraphics[width=0.5\textwidth, trim={0 0 4.9cm 0}, clip]{26_Dissertation_02_generator_comparison_64-eps-converted-to.pdf}
% 	\caption[Workflow for creating a pseudo-spectral synthetic trace.]{Workflow for creating a pseudo-spectral synthetic trace.}
% 	\label{fig:data_generators}
% \end{figure}

\subsection{DNN Framework}\label{sec:results_DNN_Framework}
\begin{figure}[h!]
	\centering
	\subfloat[Fourier components.\label{fig:DNN_framework_1}]{%
        \includegraphics[width=0.8\linewidth]{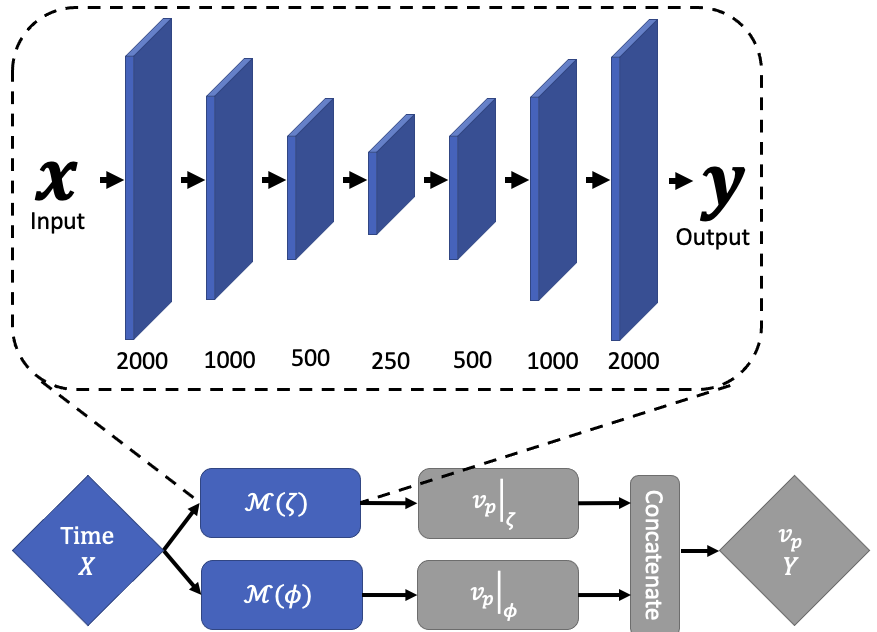}}
    \\
  \subfloat[Continuous wavelet transform.\label{fig:DNN_framework_2}]{%
        \includegraphics[width=0.8\linewidth]{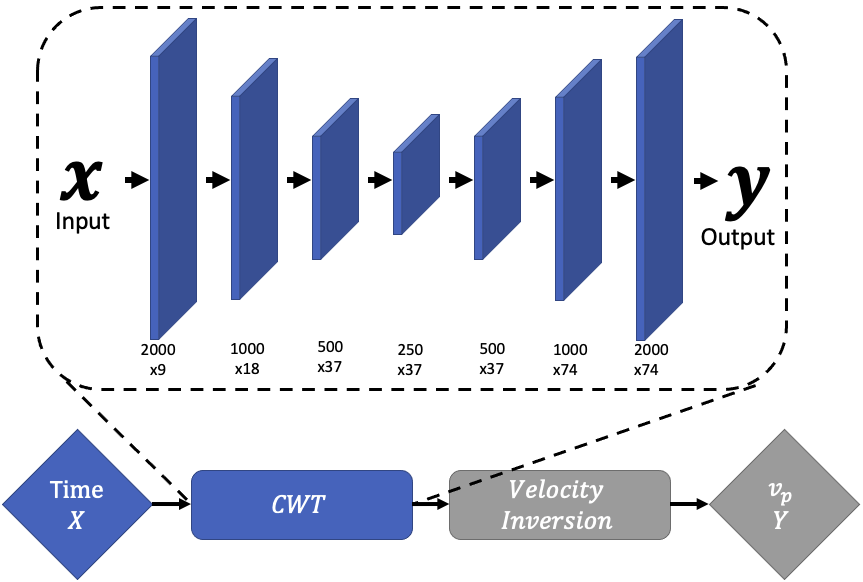}}
		
	\caption[Pseudo-spectral FWI DNN workflow.]{Pseudo-spectral FWI DNN frameworks to invert for Fourier Transform and CWT. $X$ is the input time domain, $Y$ is the output $v_p$ velocity and $\mathcal{M}$ is the Fourier domain, with magnitude $\zeta$ and phase $\phi$. Each component (blue or grey box) is a network.}         
	\label{fig:DNN_frameworks}
\end{figure}

Figure~\ref{fig:DNN_framework_1} illustrates the DNN framework used to first invert for the Fourier coefficients from the time domain and then invert for velocity profile. The complete workflow has five modules, with each module consisting of a neural network with 5 fully connected hidden layers. The layer distributions is shown within each fiure. This hour-glass design can be considered representative of multi-scale FWI \citep{Bunks1995} since at each hidden layer, the NN learns an abstracted frequency component of the data at a different scale. This is synonymous with modern DNN approaches such as encoder-decoders and U-Net \citep{ronneberger2015u} and how they extract data representations. The final concatenate network learns the optimal way for combining the outputs.
%  In total, the DNN had 25 hidden layers.
In the case of the CWT, we designed a similar framework, except that there is an additional dimension to be able to create the CWT. This is shown in Figure~\ref{fig:DNN_framework_2}. 

\subsection{Pre-Processing}\label{sec:results_Normalisation}
In classical DNN approaches, it is best practice to normalise or standardize the dataset. An experiment was executed to assess what would happen with and without normalisation of the data for the Magnitude component architecture for 10,000 training and 1,000 validation traces. 
% The computation and memory resource requirement for this test were small enough such that this was executed on a 2.6 GHz 6-Core Intel Core i7, 16GB RAM personal computer. 
The scaling approaches considered are the Standard Scaler and Min-Max Scaler, defined as:
\begin{equation}
    \text{Min-Max Scaler :} \quad x_{MM} = \frac{x-x_{MIN}}{x_{MAX}-x_{MIN}}, \\
\end{equation}
\begin{equation}
    \text{Standard Scaler :} \quad x_{SS} = \frac{x-\mu}{\sigma},
\end{equation}
where $x_{MIN}$, $x_{MAX}$ are the minimum and maximum values of the data, $\mu$ is the mean and $\sigma$ is the standard deviation of the training samples.

\begin{figure}[h]
    \centering
    \includegraphics[width=0.5\textwidth]{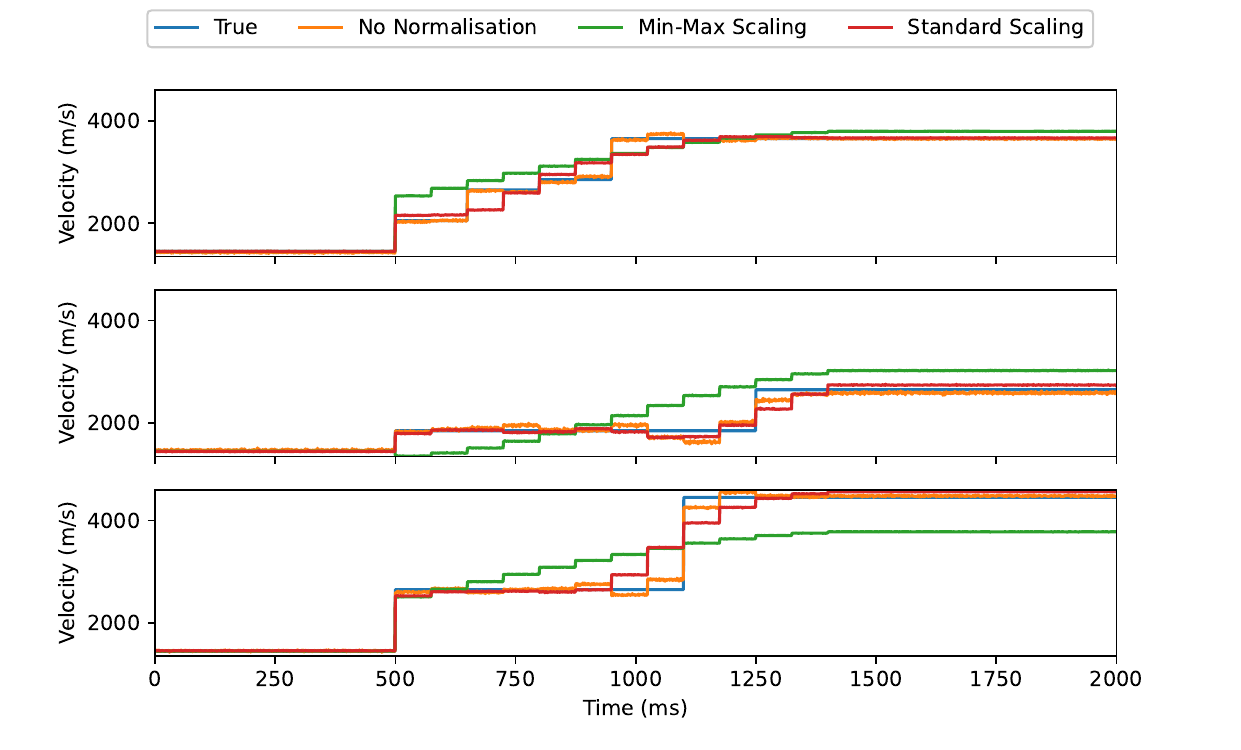}
    \caption[Comparison of velocity profiles with and without pre-processing.]{Comparison of velocity profiles with and without pre-processing.}
    \label{fig:scaling_velocity}
\end{figure}

% The mean square error for the dataset with-out and with processing was evaluated and is shown in Table~\ref{tab:normalisation_mse}. The value of the mean squared error indicates that pre-processing in the form of normalisation or standardization should not be applied to the problem dataset.
The impact of the pre-processing on the inverted velocity profiles is shown in Figure~\ref{fig:scaling_velocity}. Min-Max scaling was the worst performant, only able to reconstruct the first layer at 500\si{ms}. Standard Scaling was considered as a potential, however, considering the third velocity profile, we can see how without scaling, more of the second layer is being reconstructed. Furthermore, having to scale data would include an additional compute overhead. For these reasons, it was shown that normalization should not be applied for data-driven FWI inversion. 

% \begin{table}[ht!]
% 	\caption{Quantitative assessment on the impact of pre-precessing}\label{tab:normalisation_mse}
%     \footnotesize
%     \centering
%     \begin{tabular}{cc}\hline
% Pre-Processing & Mean Square Error \\ \hline
% No Normalisation          & 4,041 \\
% Min Max Normalisation     & 296,672 \\
% Standard Scaling          & 17,653 \\\hline
% \end{tabular}
% \end{table} 

% Figure~\ref{fig:scaling_metrics} illustrates the loss value with and without scaling for both the training and validation dataset. Looking at just these metrics for network evaluation, it is evident that scaling should be implemented as it is improving the network performance by at least order of 4. When doing the prediction and inverse-scaling on the test data, the reconstruction of the data from the scaling is of poorer quality. 

% \begin{figure}[!ht]
% 	\centering
% 	\subfloat[Comparison of DNN performance metrics with and without scaling.\label{fig:scaling_metrics}]{%
% 		\includegraphics[width=0.5\textwidth]{26_Dissertation_07_Normalization_01_metric_performance-eps-converted-to.pdf}}
% 	\\
%   	\subfloat[Comparison of velocity profiles with and without scaling.\label{fig:scaling_velocity}]{%
% 	  \includegraphics[width=0.5\textwidth]{26_Dissertation_07_Normalization_02_vel_profiles-eps-converted-to.pdf}}
% 	\caption[Comparison on the application of data scaling.]{Comparison on the application of data scaling.}\label{fig:application_scaling}
% \end{figure}

\subsection{Architecture Comparison}\label{sec:results_Architecture_Comparison}
We proceed trying to identify the ideal configuration in terms of architecture, loss, time-to-train and over-fitting.
%  The following computations were made using Python 3.7 and Tensorflow 2.0.0 with a Keras backend. These were executed on an NVIDIA Titan V Graphical Processing Unit with 5120 cores and 12GB ram provided in collaboration with Dr. Carlo Giunchi at Istituto Nazionale di Geofisica e Vulcanologia at Pisa.
The conversion from time trace to pseudo-spectral representation was fixed for all 1D and 2D networks such that the comparison was on the architectures. 
% Figure~\ref{fig:dnn_fwi_1_comparison_dnn_arch} gives a comparison of different DNN architectures, loss optimizers, duration of training and validation curves. 

The networks were trained for the same number of epochs without early stopping. The training and validation data consisted of 1,000,000 and 100,000 generated traces. The loss was fixed to be the Mean Squared Error (MSE)
%  and lr represents the learning rate on a secondary $y$-axis in Red. 
The architectures chosen for this comparison are a Multi-Layer Perceptron (MLP), Convolutional neural network with 1D filters (Conv1D) and 2D filters (Conv2D), VGG~\citep{Simonyan2014} and ResNet~\citep{He2016IEEE}. The loss optimizers were Adagrad~\citep{Duchi2011}, Adadelta~\citep{Zeiler2012}, RMSprop~\citep{Hinton2012} and Adam.

Evaluation for all the loss curves is summarised in Table~\ref{tab:quantitative_assessment_arch_loss}. The criteria below are ranked from 1-20, with 20 being the best result:
\begin{itemize}
    \itemsep0em
    \item Duration ($d$): 	Duration of training. The shorter the duration, the better the performance.
    \item Train ($t$): 		Lowest MSE within training.
    \item Validation ($v$): Qualitative assess of under-fitting/over-fitting and learning rate performance.
    \item Inversion ($i$):	RMSE of 100,000 validation velocities as compared to true velocity.
\end{itemize}
The score was calculated as
\begin{equation}
    \text{Score} = d+t+v+2i.
\end{equation} 
This formula is arbitrarily chosen and linear in nature, making ideal for interpretation and understanding. The additional weight of 2 for the inversion rank emphasizes the inversion is the most important criteria. The overall rank criteria grades all and the best performer will have the lowest overall ranl. The best performing architecture-loss combination was identified as Conv1D-Adadelta. 

% Table~\ref{tab:quantitative_assessment_summary_arch} and Table~\ref{tab:quantitative_assessment_summary_loss} summarize Table~\ref{tab:quantitative_assessment_arch_loss} per architecture and loss optimizer respectively. Table~\ref{tab:quantitative_assessment_summary_arch} further reinforces the choice for ideal setup being of type Conv1D since this architecture ranked in the top four, irrespective of Loss Optimizer. Table~\ref{tab:quantitative_assessment_summary_loss} is in agreement that Adadelta is the better loss optimizer for our setup, however the difference is relatively small and not substantial. Choosing a different loss optimizer would not result in deterioration of our result.

\begin{table}[ht!]
	\caption{Quantitative assessment for architecture and loss optimizers.}\label{tab:quantitative_assessment_arch_loss}
    \footnotesize
    \centering
    \begin{tabular}{@{}cccccccc@{}}\hline
Architecture & \makecell{\scriptsize{Loss} \\ \scriptsize{optimizer}} & $d$ & $t$ & $v$ & $i$ & Score & \makecell{\scriptsize{Overall} \\ \scriptsize{Rank}} \\ \hline
MLP          & Adagrad        & 18       & 4     & 20         & 11        & 64    & 6    \\
MLP          & Adadelta       & 16       & 13    & 19         & 8         & 64    & 6    \\
MLP          & RMSprop        & 20       & 7     & 9          & 15        & 66    & 5    \\
MLP          & Adam           & 19       & 19    & 9          & 7         & 61    & 8    \\
Conv1D       & Adagrad        & 17       & 2     & 15         & 18        & 70    & 4    \\
Conv1D       & Adadelta       & 14       & 8     & 19         & 19        & 79    & 1    \\
Conv1D       & RMSprop        & 14       & 9     & 9          & 20        & 72    & 3    \\
Conv1D       & Adam           & 15       & 10    & 17         & 17        & 76    & 2    \\
Conv2D       & Adagrad        & 12       & 5     & 15         & 12        & 56    & 9    \\
Conv2D       & Adadelta       & 12       & 15    & 15         & 4         & 50    & 14   \\
Conv2D       & RMSprop        & 10       & 1     & 9          & 14        & 48    & 15   \\
Conv2D       & Adam           & 10       & 20    & 3          & 10        & 53    & 11   \\
VGG          & Adagrad        & 8        & 3     & 15         & 13        & 52    & 13   \\
VGG          & Adadelta       & 8        & 14    & 15         & 9         & 55    & 10   \\
VGG          & RMSprop        & 1        & 11    & 9          & 16        & 53    & 11   \\
VGG          & Adam           & 8        & 12    & 4          & 5         & 34    & 17   \\
ResNet       & Adagrad        & 4        & 6     & 17         & 2         & 31    & 19   \\
ResNet       & Adadelta       & 4        & 17    & 15         & 3         & 42    & 16   \\
ResNet       & RMSprop        & 2        & 16    & 3          & 6         & 33    & 18   \\
ResNet       & Adam           & 5        & 18    & 3          & 1         & 28    & 20   \\\hline
\end{tabular}
\end{table}

% \begin{table}[!htb]
%     \centering
%     \begin{minipage}[b]{.45\textwidth}
% 		\caption{Quantitative assessment for architectures.}
%             \label{tab:quantitative_assessment_summary_arch}
%         \centering
%             \begin{tabular}{cccc}\hline
%                 \multirow{2}{*}{Architecture} & \multicolumn{3}{c}{Score} \\ \cline{2-4} 
%                                             & Avg      & Min   & Max   \\ \hline
%                 Conv1D                        & 74.3 & 70 & 79 \\
%                 MLP                           & 63.8 & 61 & 66 \\
%                 Conv2D                        & 51.8 & 48 & 56 \\
%                 VGG                           & 48.5  & 34 & 55 \\
%                 ResNet                        & 33.5  & 28 & 42 \\\hline
%             \end{tabular}
%     \end{minipage}
%     \qquad
%     \begin{minipage}[b]{.45\textwidth}
% 		\caption{Quantitative assessment for loss optimizers.}
%             \label{tab:quantitative_assessment_summary_loss}
%         \centering
%             \begin{tabular}{cccc}\hline
%             \multirow{2}{*}{Architecture} & \multicolumn{3}{c}{Score} \\ \cline{2-4} 
%                                         & Avg      & Min   & Max   \\ \hline
%             Adadelta                      & 58.0   & 42 & 79 \\
%             Adagrad                       & 54.6 & 31 & 70 \\
%             RMSprop                       & 54.4 & 33 & 72 \\
%             Adam                          & 50.4 & 28 & 76 \\\hline
%             \end{tabular}
%     \end{minipage}  
% \end{table}

\subsection{Marmousi Model Experiment}\label{sec:results_marmousi_dataset}
The Marmousi-2 model \citep{Martin2002} was used to evaluate data-driven DNN on an industry standard dataset. This was filtered with a 150m vertical median filter and the number of layers in each velocity profile was analytically calculated to range between 20 to 50 layers. 
\subsubsection{DNN FWI Generator}\label{sec:results_marm_generator}
A generator was constructed to be able to invert for our modified Marmousi model. The parameters are given in Table~\ref{tab:marm_data_generator_params}. 
% A sample of the velocity, trace and CWT generated by this generator are available in Figure~\ref{fig:marm_generator}.
\begin{table}[h]
	\caption[Marmousi data generator parameters.]{Marmousi data generator parameters.}\label{tab:marm_data_generator_params}
    \centering
    % \ra{1.3}
    \begin{tabular}{@{}lc@{}}\hline
        Description                   & Value \\ \hline
        Length of trace (ms)              & 2801  \\
        Minimum velocity (m/s)             & 1450  \\
        Maximum velocity (m/s)             & 5000  \\
        Minimum velocity separation (m/s)   & 15    \\
        Minimum time sample (ms)           & 0     \\
        Maximum time sample (ms)          & 2801  \\
        Minimum time separation (ms)      & 5     \\
        Minimum number of layers      & 20    \\
        Maximum number of layers      & 50    \\
        Dominant frequency (Hz) & 3.5     \\ \hline
    \end{tabular}
\end{table}

% \begin{figure}[h]
%     \centering
%     \includegraphics[width=0.9\textwidth]{27_real_data_07_Marmousi_Generator_Sample-eps-converted-to.pdf}
%     \caption[Sample velocity profile, trace and CWT generated by Marmousi generator.]{Sample velocity profile, trace and CWT generated by Marmousi generator.}
%     \label{fig:marm_generator}
% \end{figure}

\subsubsection{DNN Training}\label{sec:results_marm_DNN_Training_and_Architecture_Performance}
The network was trained for 30 epochs, at 1,000,000 traces and 100,000 traces per training and testing dataset respectively. The network was a slightly modified version of the ideal network identified in the previous section due to the longer time length of trace.

\subsubsection{Classical FWI}\label{sec:results_classical_FWI}
3.5\si{Hz} FWI with Sobolev space norm regularization\citep{Kazei2019} was used to compare against data-driven FWI. This results in a minimum update resolution of 414m, and the iterative update process started from frequency 1\si{Hz} and iteratively updated by a factor of 1.2. The optimization algorithm was L-BFGS-B, with 50 iterations per frequency band in each update. Forward shot modelling was done every 100m, starting from 100m offset, and receivers spaced every 100m. 

% \begin{figure}[h]
%     \centering
%     \includegraphics[width=0.78\textwidth]{27_real_data_03_FWI_results_01_vel_image-eps-converted-to.pdf}
%     \caption[Classical FWI with Sobolev space norm regularization.]{Classical FWI with Sobolev space norm regularization result. Top: Initial modified Marmousi model. Middle: Initial velocity provided for FWI. Bottom: FWI result following from inverting for 3.6\si{Hz}.}
%     \label{fig:marm_classical_fwi}
% \end{figure}

% \begin{figure}[h]
%     \centering
%     \includegraphics[width=0.99\textwidth]{27_real_data_03_FWI_results_02_vel_profile-eps-converted-to.pdf}
%     \caption[Velocity profiles through Xlines on Marmousi, Initial and FWI results.]{Velocity profiles through Xlines on Marmousi, Initial and FWI results as shown in Figure~\ref{fig:marm_classical_fwi}.}
%     \label{fig:marm_classical_fwi_velocity}
% \end{figure}

\begin{figure}[h]
    \centering
    \subfloat[Top: Initial Marmousi model with highlighted Zoom 1-3. Middle: DNN FWI result. Bottom: Classical FWI result following from inverting for 3.5 Hz.
    \label{fig:comparison_dnn_fwi_image}]{%
	\includegraphics[width=0.5\textwidth
    % , trim={0 9.3cm 0 0}, clip
    ]{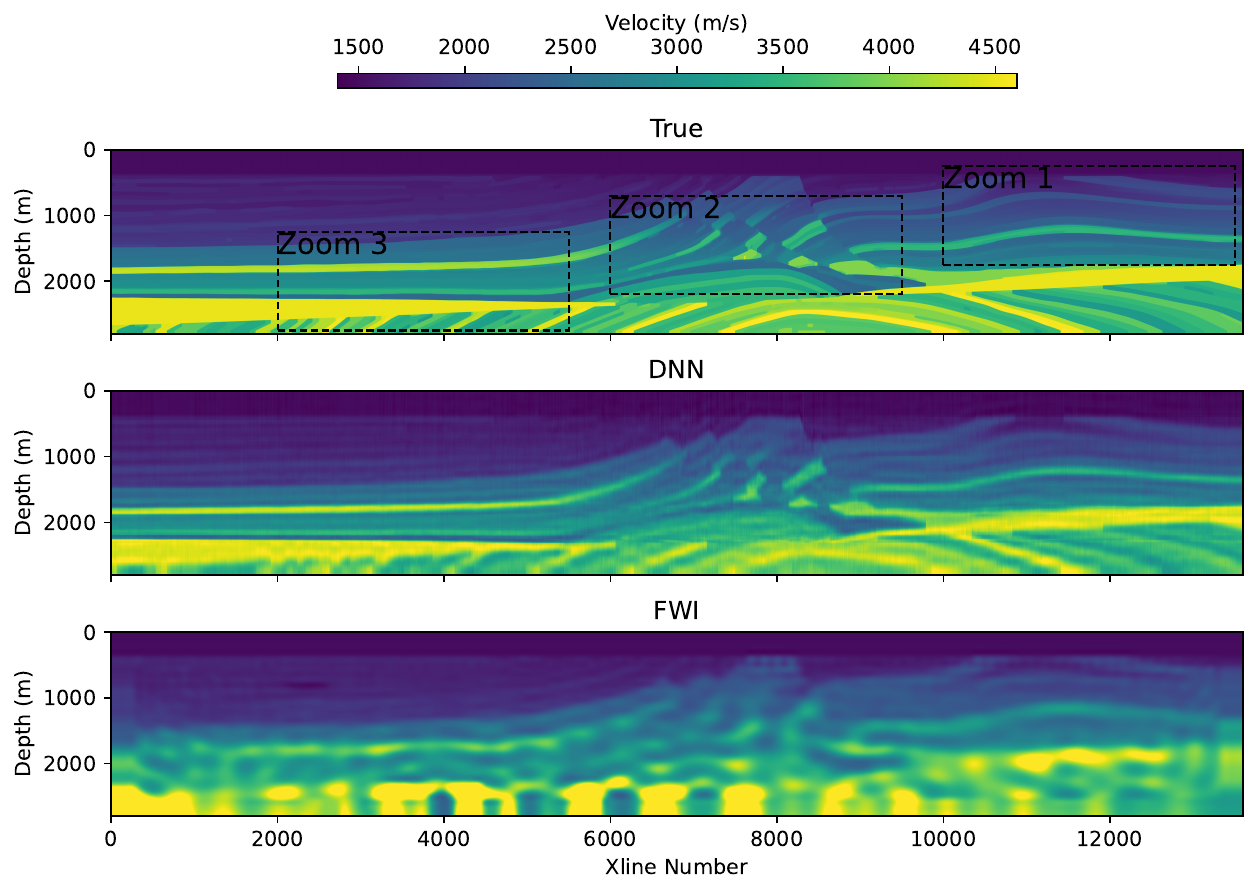}
    }
    \\
    % \subfloat[Zoomed comparison of DNN and Classical FWI velocity models and corresponding errors.
    % \label{fig:comparison_dnn_fwi_image_zoom}]{%
    % \includegraphics[width=0.5\textwidth]{27_real_data_05_DNN_FWI_COMPARISON_01_vel_image_Zoomed_In-eps-converted-to.pdf}
    % }
    % \\
    \subfloat[Xline velocity profiles.
    \label{fig:comparison_dnn_fwi_velocity}]
    {%
    \includegraphics[width=0.5\textwidth]{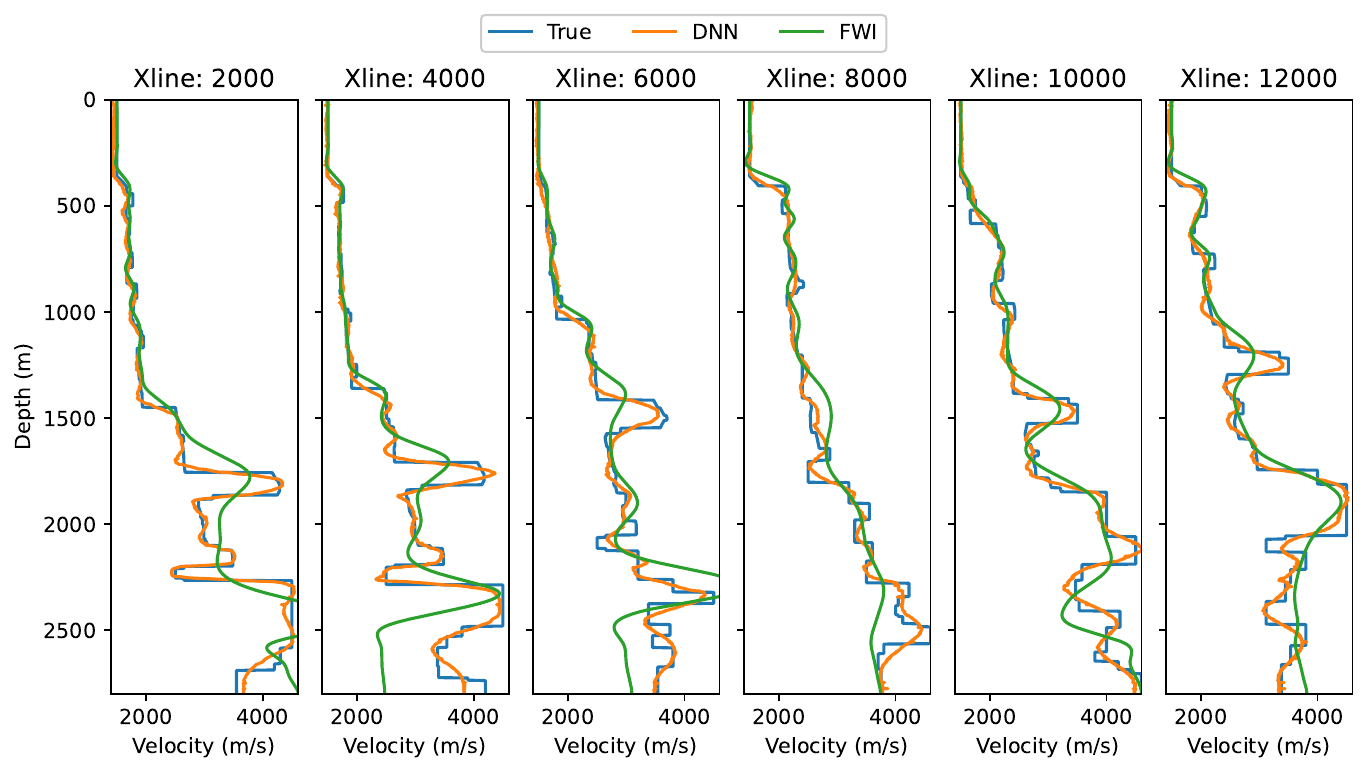}
    }
    \caption{Comparison of data-driven DNN and Classical FWI on Marmousi.}
\end{figure}

% \subsubsection{DNN and Classical FWI}\label{sec:results_comparison_dnn_fwi}
Figure~\ref{fig:comparison_dnn_fwi_image} compares classical FWI and data-driven DNN FWI. Off the start, it is clearly evident how the DNN approach is producing a lot more uplift than the standard approach. There is improved imaging in the sediment layers, with distinct layers being reconstructed which would otherwise be missed with classical FWI – Zoom 1 
% in Figure~\ref{fig:comparison_dnn_fwi_image_zoom}. 
The middle section, with the heavily over-trusted layers shown in Zoom 2, the velocity layers are also being reconstructed to good levels and the small sedimentary pockets at the pinch of the over-thrust are being to be imaged as well. These are being missed completely in Classical FWI. Sub-salt in Zoom 3, DNN is once again producing much better imaging up to the salt and below the salt. Indeed, sub-salt, we are starting to image partially some of the layer coming up into the salt. 
% The inversion process is not perfect as shown by the differences in velocities in Figure~\ref{fig:comparison_dnn_fwi_velocity}.
%  Comparison of the error maps, the problematic areas of DNN are also those for classical FWI. 

% In Zoom 1, the amplitude of the large velocity layer coming in at 1400m depth is not being inverted properly. The onset of this layer is not as problematic, but leakage is evidently present. Similarly, for Zoom 3, the salt arrival at 2200m depth, is being imaged by DNN and not by FWI. In Zoom 2, the error hotspot for DNN are similar to those of FWI, however the magnitude of the error is of an order different. This would indicate that DNN is very good performant when it comes to inverting for large velocity packages. 
% Figure~\ref{fig:comparison_dnn_fwi_spectra} gives the amplitude spectra for the full and zoomed velocity models respectively. This show that the frequency content is similar in either approach, yet both are lower than the true.

The velocity profiles (Figure~\ref{fig:comparison_dnn_fwi_velocity}) 
% and trace reconstruction (Figure~\ref{fig:comparison_dnn_fwi_trace}) 
confirm that our DNN approach inverted more of the signal than classical FWI. Upon closer investigation, we are seeing small spikes on the velocity on the salt section of Xlines 2000, 4000 and 6000. Further training would potentially mitigate this. From the velocity profiles, we see how FWI is able to update the shallow sections up to 1400m really well, potentially better than DNN as it is able to identify a velocity inversion at depth 500m on Xline 8000 and a pronounced segment layer at depth 800m on Xline 12000. However, beyond 1400m depth, the geometry and forward-modelling physical constraints from ray-tracing come into play and FWI is unable to provide more uplift at deeper velocity packages.

\section{Discussion}\label{ref:sec_disc_development_dataset}
\subsection{Inversion Paradigm}
Within the DNN approach, the inversion component is data-driven and the generator is based on geophysics. If data-generators are to include information that might not be deterministically available, the inversion process could invert for this additional information. These types of data-driven models could be pre-cursors for deterministic models \citep{Araya-Polo2018}. 

\begin{figure}[h]
    \centering
    \includegraphics[width=0.5\textwidth]{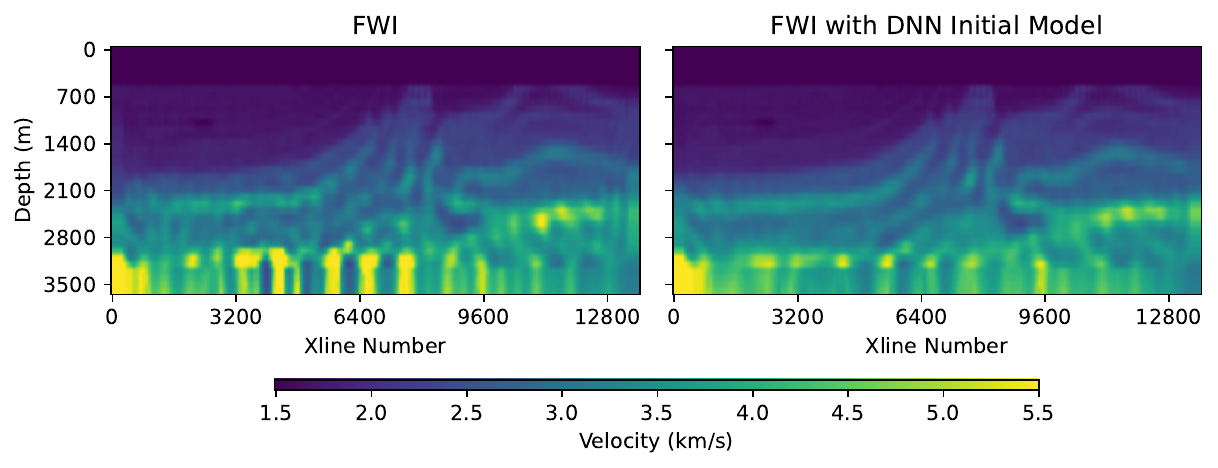}
    \caption[Improved FWI result with DNN as initial model]{Improved FWI result with DNN as initial model.}
    \label{fig:dnn_initial_fwi}
\end{figure}

Figure~\ref{fig:dnn_initial_fwi} shows inversion for classical FWI without and with DNN as a priori model. The latter approach produces more layer continuity and better imaging at depth. This comes with relatively no extra overhead cost since the DNN would be a pre-trained network.

\subsection{Data Generators for Real Data}
\begin{figure}[ht!]
\begin{centering}
	\subfloat[The first two columns are inversions for velocity profiles within the generator limit for the number of layers. For more layers, the inversion tries to generalize for these profiles but misses some components. The inversion is still able to identify some layers, but not to the same resolution as if trained on those layers.\label{fig:dnn_discussion_missing_layer}]{\includegraphics[width=0.5\textwidth]{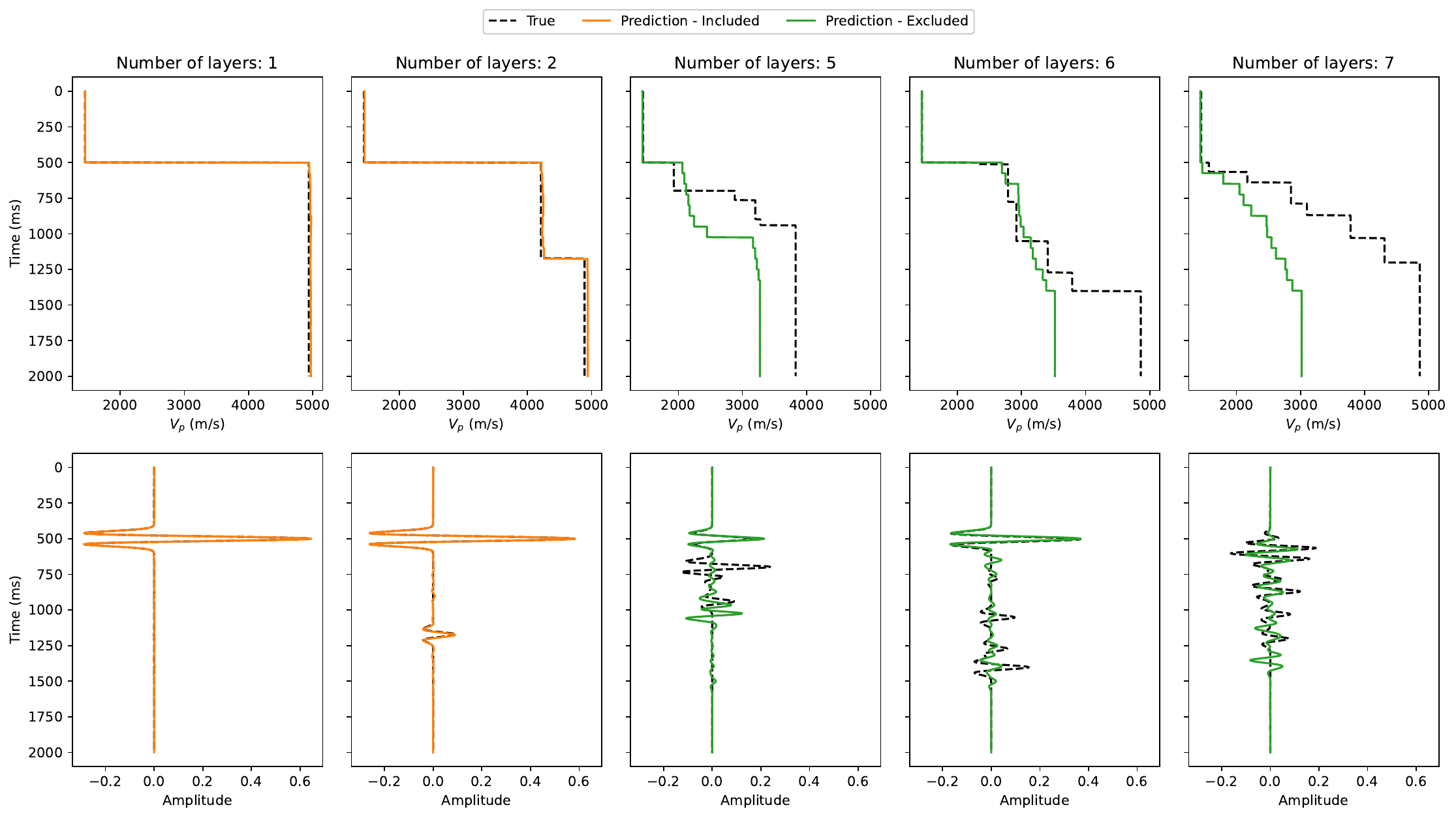}}
	\\
  	\subfloat[Large velocity inclusions and Velocity inversion would be missed as well.\label{fig:dnn_discussion_Inclusion_Inversion}]{\includegraphics[width=0.5\textwidth]{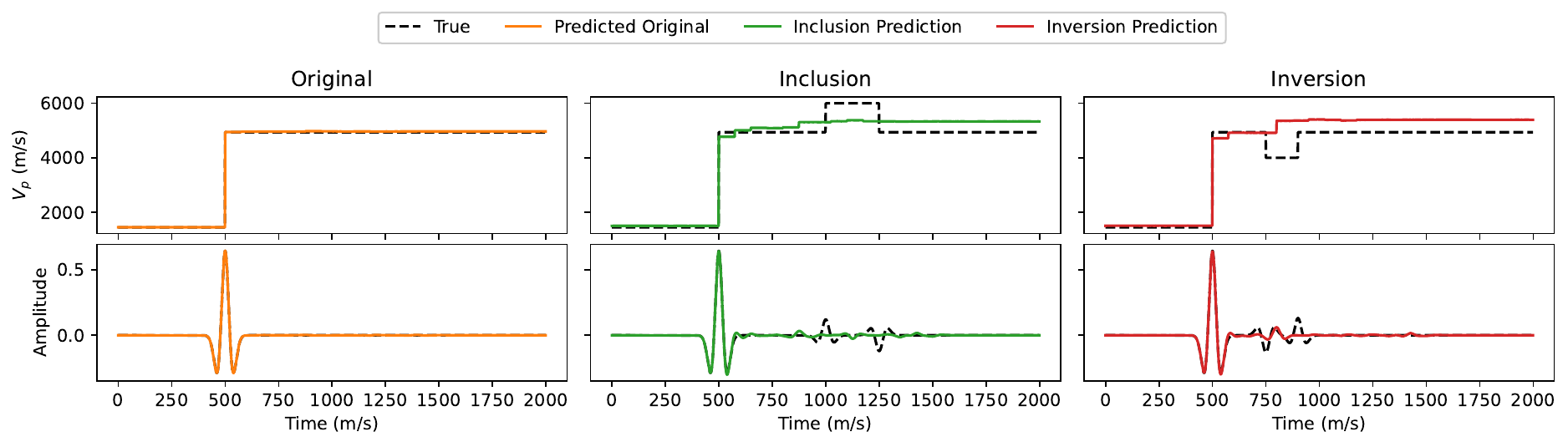}}
	\\
	\subfloat[Robustness to noise up to 25\% contamination.\label{fig:dnn_discussion_Noise_Sensitivity}]{\includegraphics[width=0.5\textwidth]{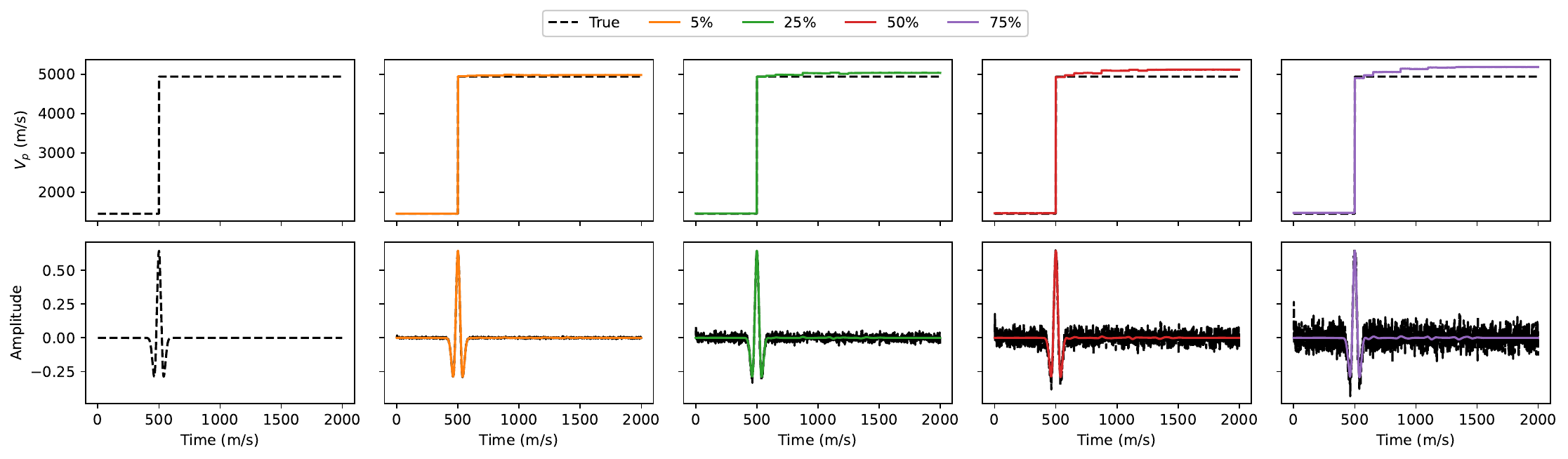}}
	\caption{The quality of the development dataset directly influence the quality of DNN inversion. A: Missing layer. B: Different geophysical models. C: Noise sensitivity. }
\end{centering}
\end{figure}

The DNN requires a global model for a real world problem. This is not always available in real world problems, however a collection of pre-trained DNN could be used as precursors of classical FWI and enable better parameterization. Consider as example a data generator trained on data that is limited to 3 layers, namely Conv1D-Adadelta DNN trained for 30 epochs for a maximum 3 layers for velocity ranging from 1450\si{ms^{-1}} to 5000\si{ms^{-1}}. If inversion is to be performed on more than 3 layers, the inversion process will start degrading as shown in Figure~\ref{fig:dnn_discussion_missing_layer}. Figure~\ref{fig:dnn_discussion_Inclusion_Inversion} illustrates inversion for a large velocity of 6000\si{ms^{-1}} and velocity inversion respectively. Given that these models were not included in the development dataset, the inversion process would never be able to invert for these velocity types correctly. 

On the other hand, the DNN framework is robust to different noise levels. Figure~\ref{fig:dnn_discussion_Noise_Sensitivity} showcases the inversion for gaussian noise contaminated data at different levels. The inversion remains relatively unaltered before 25\% and then starts degradation. This inversion process could be used as a de-noising technique given the correct data-generator.

% Modelling techniques and transform spaces could provide an alternative approach for the DNN. \citep{Jozinovic2020} use Wigner-Ville distributions for pseudo-spectral representations of the seismograms for prediction of intensity measurements of ground shaking, or other representations such as Recurrence Plots \citep{Kamphorst1987}, Markov Transition Fields \citep{Wang2015} and Gramian Angular Fields \citep{Wang2015}.

\section{Conclusion}
In this manuscript elements within a classical FWI framework were shown to be substitutable with DNN components. The base architecture for the network was set to be an hour-glass neuron design, representative of multi-scale FWI and modern DNN approaches. This was tested for normalization and concluded not applicable. MLP, 1D and 2D convolutions, VGG and ResNet type architectures for Adagrad, Adadelta, RMSprop and Adam optimizer were quantitatively evaluated for training duration, performance, validation and learning rate performance, and inversion. The best performing architecture-loss combination was identified as Conv1D-Adadelta. 
% Conv1D architecture ranked the highest in all tests, whilst the differences in optimizer were superficial. The choice of architecture was the most important aspect as choosing a different loss optimizer would not result in deterioration of the result.

The Conv1D-Adadelta network was trained for inversion of a modified Marmousi model by using a 20-50 layer generator, with velocities ranging from 1450\si{ms^{-1}} to 5000 \si{ms^{-1}}. 3.5Hz classical FWI was compared to this data-driven DNN inversion. Inversion performance in shallow sections was equally good for either classical FWI or DNN approach. DNN framework performs better for deeper and over-thrust areas since DNNs are not bound by forward-modelling physical constraints from ray-tracing.

Data-driven FWI should be considered within global approximation approaches and has potential to be used as \emph{a priori} to deterministic FWI. The DNN model generators were shown to work within the boundaries of their parameter. Application of pre-trained networks is relatively easy and thus different geophysical model hypothesis could be assessed quickly. Pseudo-spectral DNN FWI is robust to noise and future work would involve implementation as a de-noising techniques. 
% Data-driven pseudo-spectral FWI is still at the very early stages of development, and more research opportunity is still available. However, to truly assess the applicability and relevance of these frameworks, this approach will have to be applied to real data in the future.

\bibliographystyle{unsrt}  
\bibliography{references}  %%% Remove comment to use the external .bib file (using bibtex).

\end{document}